\documentclass[dvips]{arxstspdf}
\usepackage{graphics,flushend}

\volume{25}
\issue{2}
\pubyear{2010}
\firstpage{227}
\lastpage{244}
\doi{10.1214/09-STS315}

\begin{document}
\begin{frontmatter}

\title{Bayesian Models and Decision Algorithms for Complex Early
Phase Clinical Trials}
\runtitle{Bayesian Designs for Early Phase  Trials}

\begin{aug}
\author{\fnms{Peter F.} \snm{Thall}\ead[label=e1]{rex@mdanderson.org}}
\runauthor{P. F. Thall}

\affiliation{University of Texas, M. D. Anderson Cancer Center}

\address{Peter F. Thall is Professor, Department of Biostatistics,
 University of Texas, M. D. Anderson Cancer Center, Houston, Texas, USA
  \printead{e1}.}

\end{aug}

\begin{abstract}
An early phase clinical trial
is the first step in evaluating the effects in humans of a potential new anti-disease agent or combination of agents.
Usually called ``phase I'' or ``phase I/II'' trials, these experiments typically have
the nominal scientific goal of determining an acceptable dose, most often based on adverse event probabilities.
This arose from a tradition of phase I trials to evaluate cytotoxic agents for treating cancer,
although some methods may be applied in other medical settings, such as treatment of stroke
or immunological diseases.  Most modern statistical designs for early phase trials
include model-based, outcome-adaptive decision rules that choose doses for successive patient cohorts
based on data from previous patients in the trial.
Such designs have seen limited use in clinical practice, however, due to their complexity, the requirement of intensive,
computer-based data monitoring, and the medical community's resistance to change.  Still, many actual
applications of model-based outcome-adaptive designs
have been remarkably successful in terms of both patient benefit and scientific outcome.
In this paper I will review several Bayesian early phase trial designs
that were tailored to accommodate specific complexities of the treatment regime and patient outcomes
in particular clinical settings.
\end{abstract}


\end{frontmatter}

\section{Introduction}\label{sec1}

\subsection{An Early Phase Trial}

Clinical trials are much more complex than
typical statistical designs may indicate.  An example is
a phase~I stem cell transplantation (SCT) trial in which  the continual reassessment method
(CRM, O'Quigley, Pepe and Fisher, \citeyear{1990OQuigleyPe}; O'Quigley, \citeyear{1990OQuigley})
was applied to optimize the per-administration dose (PAD) of gemcitabine, $d_G,$
when added to an established two-agent preparative regimen consisting of intravenous busulfan and melphalan
(Andersson et al., \citeyear{2002Andersson}).
The design was used for each of two separate, parallel trials, one
for allogeneic  transplant (allotx), which uses stem cells from a matched donor, and one for autologous  transplant (autotx),
which uses the patient's own stem cells.
For each patient, during the period from day $-$10 to day $-$1 preceding the SCT on day 0,
each of the three agents was given on two or more days using a particular schedule and PAD.
Previously, a six-day schedule of additional gemcitabine had been tried, but it was found to be too toxic, so
in this trial each patient's assigned $d_G$ was given in a two-day schedule,
on each of days $-$8 and $-$3, for total gemcitabine dose $2d_G$.   ``Toxicity'' was defined to be any
regimen-related grade 4 or 5 adverse event (AE) occurring within 30 days post transplant and
affecting a vital organ, but excluding  AEs that occur routinely in SCT,
such as marrow suppression and, in allotx, graft-versus-host disease (GVHD).  Using the usual CRM criterion,
the design's nominal goal was to find $d_G$ from a predetermined  set of 10 PADs ranging from 225 to 3675
mg$/$m$^2$ having toxicity probability, $\pi(d_G,\theta),$ with posterior mean  $\mathrm{E}_{\theta}\{\pi(d_G,{\theta})|\mathit{data}\}$
closest to the target 0.10, where $\theta$ denotes the model parameters. The principal investigator (PI) specified the conservatively low
target 0.10 in part due to the previous negative experience with the six-day schedule, and also because 0.10 is consistent with
the toxicity rate of the established two-agent regimen. In each subgroup, gemcitabine doses were to be chosen for successive cohorts of 3 patients,
up to a maximum of 36 patients, with the safety rules that no untried dose could be skipped when escalating and
accrual to the subgroup would be stopped if the lowest dose $d_G = 255$ was unacceptably toxic,
formally if $\operatorname{Pr}\{\pi(225,{\theta})> 0.10 | \mathit{data}\} > 0.80$.

Because clinical trials are medical experiments with human subjects, they
often do not play out precisely as designed.  In the course of this trial: (i) when no toxicity was seen
in the first 24 patients the PI decided to change the two-day
gemcitabine schedule ($-8, -3$) to the three-day schedule ($-8, -6, -3$) while maintaining the same total dose by giving
$\frac{2}{3} d_G$ on each day, (ii)~this three-day schedule was quickly found to cause severe skin toxicity in the first few patients
who received it and accrual was suspended, and (iii) we re-designed the trial again by returning to the two-day schedule, but
(iv) at the PI's request we also expanded the set of possible $d_G$ values.  After 28 allotx patients had been treated and
fully evaluated, however, (v) concern about observed grade 3 mucositis and skin toxicities seen at higher dose levels
caused the physicians to expand the definition of ``toxicity'' to include these events, which previously had been
excluded if they could be resolved therapeutically within two weeks.  Along with this change, they also decided to
change the CRM target from 0.10 to 0.15.   These last changes had the combined effect of substantially
increasing the numbers of patients with ``toxicity'' among those
treated at higher dose levels and greatly
reducing the value of $d_G$ recommended by the CRM.  Per standard regulatory procedure,
it was necessary to obtain institutional review board approval  for each change in the design.
So, this trial actually was designed five times,
it evaluated effects of a combination of three agents given in overlapping pre-transplant schedules,
both the dose and schedule of gemcitabine were varied adaptively during the trial using three different formulations of
the CRM to choose gemcitabine PADs and {ad hoc} decisions for changing schedules,
there were two simultaneous trials involving different SCT modalities,
the dominant effects on toxicity were both $d_G$ and the schedule of gemcitabine, and
the definition of toxicity was changed
near the end of the trial to be more inclusive and thus obtain a more protective dose selection criterion.
Overshadowing all of this were the actual goals, which were not only to control toxicity but also to
reduce the rate and severity of GVHD in the allotx patients
and to improve the rates of engraftment and 100-day survival, compared to the established preparative regimen.

\subsection{Some Generalities}

 Denote the treatment administered to a given patient by $x$.
In the designs discussed here, $x$ will be the dose of an agent, the dose pair of two agents given together,
or a (schedule, dose)  combination consisting of a finite sequence of administration times and corresponding doses.
Actual patient outcome in oncology trials is very complex, often including numerous
different types of toxicity scored on ordinal scales of severity (grade), disease status
scored as a binary or ordinal variable, with each often recorded at several successive evaluations, as well as the times of
delay or discontinuation of treatment, drop-out or death.  In sharp contrast,
the outcome $Y$  used for statistical decision-making during the trial usually is defined to be a single variable or
possibly a~vector of two variables.  In the examples given here, the designs assume that
$Y$ is, respectively, a single binary toxicity indicator, a~vector of two binary indicators of toxicity and efficacy,
a vector of ordinal toxicities or a time-to-toxicity variable subject to right censoring.  The model consists of a probability
 density function (p.d.f.) or mass function (p.m.f.)  $f(y|x,\theta)$ of $Y$ for a patient who receives treatment $x$,
and a prior $p(\theta|\xi)$, where $\theta$ is the model parameter vector and $\xi$ are fixed hyperparameters.
The data observed from the first $n$ patients in the trial
are $\mathcal{D}_n = \{(x^{(1)},Y^{(1)}),\ldots,(x^{(n)},Y^{(n)})\},$ with likelihood
$\mathcal{L}_n(\mathcal{D}_n | \theta) = \prod_{i=1}^n f(Y^{(i)}|x^{(i)},\theta)$
and posterior $p_n(\theta | \mathcal{D}_n,\xi)\propto \mathcal{L}_n(\mathcal{D}_n | \theta)p(\theta|\xi).$

All of the designs that I will discuss here utilize Bayesian  ``learn-as-you-go'' decision rules
to choose $x$ from the set of possible treatments, $\mathcal{X}$,
based on the posterior $p_n(\theta | \mathcal{D}_n)$ computed from the most recent data
available when a new patient is enrolled.
Such a sequentially adaptive decision algorithm may be expressed as a sequence $\alpha = \{\alpha_n\}$
of functions $\alpha_n\dvtx \mathcal{D}_n \rightarrow \mathcal{X}\cup \phi,$
where  $\phi$ denotes the empty set and $\alpha_n(\mathcal{D}_n) = \phi$ means
``Do not treat with any $x \in \mathcal{X}$.''
In general, $\{\alpha_n\}$ may include several adaptive decision rules
used together, such as rules for choosing a dose, a dose pair or a (schedule, dose) combination,
for temporarily suspending accrual to wait for additional data on previously treated patients,
or for stopping the trial early because no $x\in \mathcal{X}$ is acceptable.
The $(n+1)${st} iteration of the
Bayesian medical decision-making process may be described by the sequence of mappings
\begin{eqnarray}\label{seq}
\quad \mathcal{D}_n &\stackrel{(f,p)}{\longrightarrow}& p_n(\theta | \mathcal{D}_n, \xi)
\stackrel{\alpha_n}{\longrightarrow}  x_{n+1}\nonumber
\\[-8pt]\\[-8pt]
&\longrightarrow& Y_{n+1} \longrightarrow  \mathcal{D}_{n+1}\nonumber
\end{eqnarray}
in which Bayes' Theorem uses the assumed probability model $(f,p)$ to map the observed data into a posterior,
the decision rules $\alpha_n$ use this to choose the next treatment $x_{n+1}$,
the patient is treated, the outcome $Y_{n+1}$ is observed, and $(x_{n+1},Y_{n+1})$ are incorporated into the data.
This process is repeated until the end of the  trial, which may be when either
a maximum sample size $N_{\mathit{max}}$ or trial duration $T_{\mathit{max}}$ is reached, or because the trial is stopped early.
The process whereby the expanding data set is used by applying Bayes' Theorem repeatedly
to turn $p_n(\theta | \mathcal{D}_n,\xi)$ into $p_{n+1}(\theta | \mathcal{D}_{n+1},\xi)$ may be called ``iterative Bayesian
learning,''
in that one learns about $\theta$ as additional data are observed during the trial. In the sequel,
to simplify notation I will suppress dependence of the posterior on the prior hyperparameters $\xi$.

A design consists of the  trial's entry criteria, treatments $\mathcal{X},$
set of possible patient outcomes, probability model $(f, p)$,  decision rules $\alpha,$
$N_{\mathit{max}}$ or $T_{\mathit{max}}$, and possibly a cohort size, $c$.
Since  $\alpha_n$ acts on $\mathcal{D}_n$ indirectly through $p_n(\theta | \mathcal{D}_n)$ in Bayesian adaptive designs,
evaluation of a design's properties must account for the fact that $\{p_n(\theta | \mathcal{D}_n)\}$
is a sequence of statistics.  The complexity of the process summarized by ($\ref{seq}$), even
for binary $Y$ and a single dose $x$, has motivated the routine use of computer simulation
under each of a set of assumed ``true'' $f$ as a tool to evaluate the
frequentist operating characteristics (OCs) of the design for various $\alpha$'s.
This is used as a basis for choosing decision rules, calibrating design parameters, and possibly
calibrating the prior. There is nothing ``non-Bayesian'' about using frequentist OCs
of a Bayesian design to adjust the prior and design parameters.
On the contrary, because simulating a trial that is based on a
Bayesian design allows the physician to better understand the consequences of
particular prior values, simulation provides a tool for the physician to modify his/her
prior so that it more accurately reflects what the physician actually believes.
It also is important to examine the prior's properties
in the natural parameter domain, such as the probability of toxicity at dose $x$,
$\pi(x,\theta)$, rather than in terms of elements of $\theta$ that may have
no intuitive meaning to a physician.
One should also examine the first few decisions $\alpha_n$ for each of several possible
configurations of data, in order to avoid a prior that does not make sense.
This is especially important for evaluating decisions that must be made early in the trial based on very little data,
such as choosing the second cohort's dose based on data from the first cohort of three patients.
The prior always has consequences in an early phase trial, regardless of how
``uninformative'' it may appear to be.

If one does not wish to use simulation as a design tool,
the most common alternative approach is to first specify a formal optimality criterion
and solve for $\alpha$ mathematically (cf. Haines, Perevozskaya and Rosenberger, \citeyear{2003Haines};
Dette et al., \citeyear{2008Dette}).  However, the simulation-based OCs of a design
obtained using a particular optimality criterion are often surprising, essentially because
such a design's properties are a consequence of the optimality criterion used.
One may \mbox{maximize} information, minimize the variance of a particular estimated quantity,
minimize mean or maximum sample size, control false positive or other incorrect decision probabilities,
minimize expected financial costs, minimize expected trial duration,
optimize outcomes for patients in the trial or for future patients, etc.
Since, unavoidably, such goals often are at odds with each other,  use of
the word ``optimal'' without qualification may be very misleading.

\subsection{Some Practical Issues}

Actual clinical trial logistics can be quite  complex.
While the CRM adaptively chooses a new dose from a continuum for each new patient,
Goodman, Zahurak and Piantadosi (\citeyear{1995Goodman}) proposed the practical modifications of choosing doses for successive
cohorts of several patients and limiting doses to a finite set.
In most of the outcome adaptive dose-finding
applications that I have seen, each newly chosen $x$ is given to a cohort.
Moreover, a ``do not skip'' safety rule often is imposed that
does not allow an untried dose to be skipped when escalating.
While limiting doses to a finite set of discrete values usually does not allow the
exact MTD to be chosen, the difference between the chosen dose and the true
MTD may be small, provided that the chosen sizes doses are reasonable.
For example, using the dose set \{100, 200, 400, 800, 1600\} must miss an actual MTD of 600
by at least 200, which is larger than the difference between the first two doses.
Moreover,  escalation from 400 to 800 or from 800 to 1600
may be unsafe, regardless of whether a do-not-skip rule is imposed.

For example, if a cohort size $c = 3$ is used, then
$\alpha_n$ chooses $x_{n+1} = x_{n+2} = x_{n+3}$
and, formally, $Y_{n+1},\break Y_{n+2}$ and $Y_{n+3}$ all
must be observed before updating the data and making the next decision.
However, if $Y_{n+1}$ has been fully evaluated at
the time patient $n+2$ is accrued, then $x_{n+2}$ may be chosen more reliably
by applying the decision criterion based on the updated posterior
incorporating the data $({x_{n+1},Y_{n+1}})$ from patient $n+1$; similarly, if  $Y_{n+1}$ and $Y_{n+2}$
are known when patient $n+3$ is accrued, their data may be included to choose $x_{n+3}$.
A simple approach that works surprisingly well is to use the ``look ahead'' rule:
If the possible outcomes of treated  patients for whom $Y$ has not yet been fully evaluated will not alter the
chosen~$x$ for the next patient, then treat the next patient with~$x$ without delay
(Thall et al., \citeyear{1999Thall}).
This is closely related to the general fact that the time window required to evaluate $Y$ per its definition
and the accrual rate together play critically important roles in trial conduct.
For example, if $Y = \mathrm{I}$ (toxicity within 3 months from start of therapy) and the accrual rate is 6 patients per month,
then any outcome-adaptive rule based on $\operatorname{Pr}(Y=1|x,\theta)$ is virtually
useless, since a large number of patients will be treated before the rule may be applied.
Some possible ways to implement an outcome-adaptive design in such settings
are as follows: (i) use $c = 1$ but enroll only a very small proportion of eligible
patients in the trial, (ii) use $c = 3$ or larger with accrual suspended between cohorts, but use the
look-ahead rule to improve logistical feasibility, or (iii) redefine the outcome to be time-to-toxicity,
but use a safety rule that may delay accrual interimly to allow the data
from previously treated patients to mature (cf. Bekele et al., \citeyear{2008Bekele}).

At the start of the trial, when $n = 0$,  the first treatment $x_1$ may be chosen
by applying the decision rule $\alpha_0$ based on the prior $p(\theta|\xi)$.
Methods for choosing a starting dose have  been proposed by Goodman,  Zahurak and Piantadosi (\citeyear{1995Goodman})
and Cheung (\citeyear{2005Cheung}).
Most commonly, $x_1$ is chosen by the physician based on the nature of $x$, the definition of $Y$,
the trial's entry criteria and clinical experience treating the disease. For example, a trial enrolling
prostate cancer patients with a life expectancy of six years is very different from a trial
enrolling brain tumor patients with a life expectancy of six months.
Similarly, depending on the trial's entry criteria and treatment, ``toxicity'' may be defined as anything
from severe fatigue to regimen-related death. It thus makes sense, during the prior elicitation process,
to calibrate $p(\theta)$ so that $\alpha_0$ agrees with the physician's $x_1$,
since the motivation for choosing a particular $x_1$ is based on prior experience.

In this paper I will review several designs that focus on the problem of reflecting
more fully particular complexities of $(x,Y)$.  Each design addresses some, but not all, of the
issues in the SCT trial described earlier.  These methods were motivated by problems that I~have encountered
during the process of designing early phase trials over the past 19 years.
Each design was developed by a collaborative team
including one or more physicians, one or more statisticians and a computer programmer.
Each may be called a ``phase~I'' design in that dose-finding is based
on toxicity, or a ``phase I/II'' design in that dose-finding is based
on both efficacy and toxicity, with the exception of the design described in Section~\ref{sec5} that
jointly optimizes schedule and dose (Braun et al., \citeyear{2007Braun}).
While it is tempting to think that a ``one-size-fits-all''
design encompassing all phase I/II possibilities may be constructed,
in my experience clinical research is far too complex to do this,
and each new trial design problem often has unique aspects that require a new model or method.
The particular data structure, probability model and decision rules that should be used to design a clinical trial
are best determined through careful discussion with the physicians planning the trial, and
must strike a compromise between the desire to accurately reflect the medical process and address scientific goals
while accommodating the practical realities of trial conduct.

I will not discuss methods for eliciting and calibrating priors, since this topic
could easily fill an additional manuscript.  I will not explore the ethical aspects of
adaptive decision rules either, since they also are quite complex (cf. Palmer, \citeyear{2002Palmer}).
Early phase trial design and conduct are difficult and complicated in large part due to
the tension between optimizing the benefit and safety of patients treated in the trial, and
learning about the effects of each $x$ on $Y$ to benefit future patients, as well
as economic constraints and regulatory requirements. In this regard,
a statistician constructing an adaptive design
should be mindful of the ethical issues regarding what happens, for example, to patient number 7
because (s)he was treated with $x_6$ based on how $\alpha_6$ acted on $p_{6}(\theta | \mathcal{D}_{6}).$

\section{Dose-Finding for Two-Agent Combinations}\label{sec2}

\subsection{Outcomes and Models}

Thall et al. (TMML, \citeyear{2003Thall}) proposed a method
for determining one or more acceptable dose pairs $x
= (x_1, x_2)$ of two cytotoxic agents given together, based on
a binary indicator $Y$ of toxicity.
To stabilize the model numerically, each dose is standardized so that $0 \leq x_1,\ x_2 \leq 1,$
for example, by dividing each raw dose by some maximum value, so that each $x \in\mathcal{X} = [0, 1]^2$.
The probability  model for toxicity is
\begin{eqnarray}\label{pdefpair}
&&\operatorname{Pr}(Y=1|x,\theta)\nonumber
\\
&&\quad\equiv \pi(x,  \theta)
\\
&&\quad=  \frac
{ \alpha_1 x_1^{ \beta_1} + \alpha_2 x_2^{ \beta_2}
+ \alpha_3 x_1^{ \beta_1\beta_3}x_2^{ \beta_2\beta_3} }
{1 + \alpha_1 x_1^{ \beta_1} + \alpha_2 x_2^{ \beta_2}
+ \alpha_3 x_1^{ \beta_1\beta_3}x_2^{ \beta_2\beta_3} },\nonumber
\end{eqnarray}
where $\theta = (\alpha_1,\beta_1,\alpha_2,\beta_2,\alpha_3,\beta_3).$  All elements
of $\theta$ are positive valued, which ensures that $\pi(x,  \theta)$ is a probability
and that it is  increasing in each entry of $x$.
Denoting the subvectors $\theta_j = (\alpha_j,\beta_j)$
for $j=1,2,3,$ so that $\theta = (\theta_1,\theta_2,\theta_3),$
the model (\ref{pdefpair}) contains the submodels
\begin{eqnarray}\label{sub1}
\pi_1(x_1,\theta_1) &=& \operatorname{Pr}\{Y=1 | x = (x_1,0),\theta_1\}\nonumber
\\[-8pt]\\[-8pt]
&=& \frac{\alpha_1 x_1^{ \beta_1}}{1+\alpha_1 x_1^{ \beta_1}},\nonumber
\end{eqnarray}
which is the probability of toxicity when agent 1 is given alone at dose $x_1$ and, similarly,
\begin{eqnarray}\label{sub2}
\pi_2(x_2,\theta_2) &=& \operatorname{Pr}\{Y=1 | x = (0,x_2),\theta_2\}\nonumber
\\[-8pt]\\[-8pt]
&=& \frac{\alpha_2 x_2^{ \beta_2}}{1+\alpha_2 x_2^{ \beta_2}}\nonumber
\end{eqnarray}
for agent 2 given alone at dose $x_2$.  Since $\alpha_j x_j^{ \beta_j} = \exp\{\log(\alpha_j) + \beta\log(x_j)\}$,
(\ref{sub1})  and (\ref{sub2})  are logistic models in a log standardized dose.
TMML assume that there is clinical experience with each single agent when used alone,
since this often is a requirement before investigating a combination in humans.
Since $\theta_j$ parameterizes $\pi_j(x_j,\theta_j)$ for $j=1,2$ and
$\theta_3$ parameterizes interaction  between the two agents,  a
key element of TMML's approach is that the priors $p(\theta_1|\xi_1)$ and  $p(\theta_2|\xi_2)$
are informative while $p(\theta_3|\xi_3)$ is vague. Assuming gamma priors on the elements of $\theta$
for tractability, TMML provide a detailed algorithm for eliciting
$p(\theta_1|\xi_1)$ and  $p(\theta_2|\xi_2)$, although if historical data are available, the posteriors from preliminary fits
of such data may be used as these priors for trial design and conduct.
Considering $\pi(x,  \theta)$  geometrically as a response surface over the domain $[0, 1]^2$,
this says that there is substantial prior knowledge about each of the two lines  $\{x\dvtx x_2=0\}$
and $\{x\dvtx  x_1=0\}$ on the edges of the response surface,
but otherwise little is known about the surface, so it is like a sheet
tied down at two edges but otherwise varying freely.
In particular, the meaning of $\theta_1$ in the model $\pi(x,  \theta)
= \pi(x,  (\theta_1,\theta_2,\theta_3))$
is very different from its meaning in the submodel $\pi_1(x_1,  \theta_1)$.
This is underscored by the prior effective sample sizes computed by
Morita, Thall and Mueller (\citeyear{2008Morita}) for the gamma priors given by TMML (\citeyear{2003Thall}, Section 3),
which are 547.3 for $p_1(\theta_1|\xi_1)$,
756.8 for $p_2(\theta_2|\xi_2)$, 0.01 for $p_3(\theta_3|\xi_3)$ and 1.5 for $p(\theta|\xi)$.
This says that, with respect to toxicity,  {a priori}
a lot is known about how each agent behaves when used alone, but almost nothing is known
about how the two agents behave together.

\begin{figure}[b]

\includegraphics{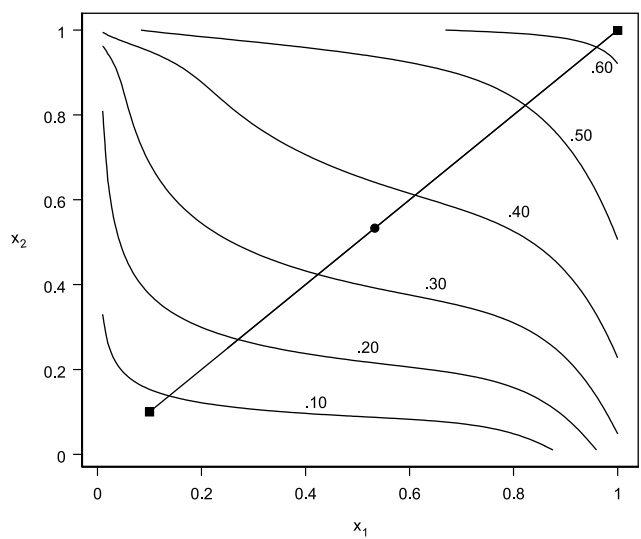}

  \caption{Isocontours of a dose-toxicity probability surface for two dimensional dose
$x = (x_1,x_2)$.}\label{fig1}
\end{figure}

\subsection{Decision Criteria}

The dose-finding method exploits the following geometric structure on $\mathcal{X} = [0, 1]^2$.
For each $p\in (0, 1)$, the set $\mathcal{X}_p(\theta) = \{x\dvtx \pi(x,  \theta) = p\}$ is the  isocontour
of all dose pairs having toxicity probability $p$.
Several isocontours for a particular fixed $\theta$  are illustrated in Figure~\ref{fig1}.
Since  $\mathcal{X}_p(\theta)\cap\mathcal{X}_q(\theta) = \phi$ if\vspace*{1pt} $p\neq q$
and $\bigcup_{0 \leq p \leq 1} \mathcal{X}_p(\theta) = [0, 1]^2$, every pair $x$
falls on a unique $\mathcal{X}_p(\theta)$ for some $p.$
The interaction term $\alpha_3 x_1^{ \beta_1\beta_3}x_2^{ \beta_2\beta_3}$ in ($\ref{pdefpair}$)  is used
instead of the simpler term $\alpha_3 x_1x_2$ in order to give the model sufficient flexibility to
allow ``S'' shaped isocontours, as shown in Figure~\ref{fig1}.

The design proceeds in two stages.
In stage 1, doses are chosen for successive cohorts from a finite set of values on the predetermined fixed diagonal line
$L_1,$  shown as the straight line at approximately 45$^{\circ}$ in Figure~\ref{fig1}.
The design is robust to the particular angle of $L_1$, as long as it is not too far from 45$^{\circ}$.
Since the response surface $\pi(x,  \theta)$ increases in each argument $x_1$ and $x_2$,
$\pi(x,  \theta)$ must increase as $x$ moves up $L_1$ from lower left to upper right.
Given target toxicity probability $\pi^*$,  dose-finding in stage 1 proceeds using the CRM criterion
of choosing $x$ for each cohort from the set on $L_1$ to minimize $|\mathrm{E}\{\pi(x,\theta)|\mathcal{D}_n \} - \pi^*|,$
starting at the lowest dose in the set, not skipping untried doses when escalating,
and adding additional doses to the set once the first toxicity is observed.
That is, restricting $x$ to $L_1$ in stage 1 reduces dose selection to a one-dimensional problem,
and a conventional CRM algorithm may be applied.
Geometrically, one may think of stage 1 as walking up and down the toxicity surface, along $L_1$,
looking for dose pairs with toxicity probability close to~$\pi^*$.

In stage 2, $x$ is chosen for successive cohorts from  the random isocontour
\begin{equation}
\mathcal{X}_{\pi^*}(\mathcal{D}_n) =   \bigl\{x\dvtx  E\{\pi(x,  \theta)| \mathcal{D}_n\}  = \pi^*\bigr\}.
\end{equation}
Since $\mathcal{X}_{\pi^*}(\mathcal{D}_n)$ contains infinitely many $x$, an additional criterion is needed
to choose one $x$ for each cohort in stage 2.  TMML suggest two criteria,
one based on the clinical criterion of ``cancer killing potential''
and the other the more usual statistical goal of maximizing Fisher Information.
Denoting the elements of $\theta$ by $\theta_1,\ldots,\theta_6$ for convenience,
the Fisher Information matrix $I(x,\theta)$ for dose $x$ has
$(j,k)$ entry
$\{\partial \pi(x,\theta)/\partial \theta_j\}\{\partial \pi(x,\theta)/\partial \theta_k\}
/[ \pi(x,\theta)\{1-\pi(x,\theta)\}]$.
Under the Bayesian model, $x$ is chosen to maximize
the posterior mean log determinant of $I(x,\theta)$ given the current data,
$\mathrm{E} [ \log \{\det I(x,\theta)\} \vert \mathcal{D}_n ]$.
Doses are chosen for successive cohorts in stage 2 by alternating between the two subsets of
 $\mathcal{X}_{\pi^*}(\mathcal{D}_n)$ to the left and right  of $L_1$.  For each subset,
the $x$ optimizing cancer killing is determined,
 the $x$ maximizing Fisher Information is determined, the average
 of these two dose pairs is computed, and the $x\in \mathcal{X}_{\pi^*}(\mathcal{D}_n)$ closest
to this average is assigned to the cohort.
 At the end of the trial, any $x\in \mathcal{X}_{\pi^*}(\mathcal{D}_n)$ is a solution.  Thus, for example,
one may  choose three final  dose pairs on $\mathcal{X}_{\pi^*}(\mathcal{D}_N)$, one on $L_1$, one to the left of $L_1$
and one to the right of $L_1,$ and randomize
patients among these three $x$ pairs in a subsequent phase II trial.

In their illustrative application, TMML use a cohort size of 2 with 60 patients
divided into $N_1=  20$ (10 cohorts) in stage 1 and $N_2 =   40$ (20 cohorts) in stage 2.
Simulations show that using $(N_1,N_2)$  either (30, 30) or (40, 20)
gives a design with inferior properties compared to (20, 40).
Cohort-by-cohort computations show that the target isocontour $\mathcal{X}_{\pi^*}(\mathcal{D}_n)$
varies substantially with each new cohort's data even after $n = 30$ or 40 patients,
but stabilizes by $n = 50$ or 60. This is the case essentially because a binary outcome
is a very small amount of information per patient.
For total sample size $N = N_1+N_2 = 60$, however, the method is quite
reliable in terms of choosing dose pairs that have true toxicity probability close to $\pi^*$.

\section{Using Both Efficacy and Toxicity}

From a clinical perspective, the primary purpose of treatment is
to fight disease, and safety is never a secondary concern in any medical setting.
Thus, both scientifically and medically,
both efficacy and toxicity matter at all stages of clinical investigation.
In many dose-finding trials,
the entry criteria specify patients with such poor prognosis that response is very unlikely,
between 4\% and 10\%.  In such settings,
targeting even a low response rate $\pi_{R^*} = 0.10$ or 0.20 and using a phase II type rule
to stop accrual if the observed response rate is likely to be below $\pi_{R^*}$ at any acceptable dose
may be impractical, since few or no responses are expected. This is the most common rationale
for conducting phase I based on toxicity alone, while recording data on biological effects
and possible clinical anti-disease effects.
However, actual response rates  vary widely between phase I trials, and
many have complete or partial response rates well over 20\% (cf. Horstmann et al., \citeyear{2005Horstmann}).
Moreover, patients enroll in a phase I trial motivated by the hope that the new treatment will
achieve an anti-disease effect, not simply the desire that no toxicity will occur.

\subsection{Outcomes and Models}\label{sec3.1}

These considerations lead to the idea that, when it is realistic to target a response rate of 10\% or larger,
dose-finding should be done using a phase I/II design based on both $E$ =  efficacy and $T$ =  toxicity.
Many phase I/II designs have been proposed
(Gooley et al., \citeyear{1994Gooley}; O'Quigley, Fenton and Hughes, \citeyear{2001OQuigley}; Braun, \citeyear{2002Braun}; Ivanova, \citeyear{2003Ivanova}).
The following  phase~I/II methodology, ``EffTox,''  is based on the developments given by
Thall and Russell (\citeyear{1998Thall}) and Thall and Cook (\citeyear{2004Thall}).
Illustrations are given by  Thall, Cook and Estey (\citeyear{2006Thall})
and Whelan et al. (\citeyear{2008Whelan}).  Patient outcome may be either a three-category
or bivariate binary variable. The former case applies when
$E$  and $T$ are defined in such a way that they are disjoint but $E \neq T^c$,
so that $Y$ takes on values in $\{E,T,N\}$  where
$N = (E\cup T)^c = {}$\{no response and no toxicity\}.
This is appropriate if, for example, toxicity is
irreversible organ damage or regimen-related death.
When it is \mbox{possible} for both $E$  and $T$ to occur, the outcome is bivariate
binary, $Y = (Y_E,Y_T)$, where $Y_k$ indicates the outcome $k=E,T.$ For either case,
denote the outcome probabilities for a patient given dose $x$ by
$\pi(x,\theta) = (\pi_E(x,\theta),\pi_T(x,\theta))$.

For the trinary outcome case, the three-parameter model used by Thall and Russell (\citeyear{1998Thall})
is  motivated by the idea that the three outcomes are ordered in the sense that $N < E < T$,
with the idea that higher $x$ is more likely to push the patient's outcome upward along this scale.
The model is given by $\pi_T(x,\theta) = g^{-1}(\beta_0 + \beta_2 x)$ and
$\pi_E(x,\theta) = g^{-1}(\beta_0 + \beta_1  + \beta_2 x) - \pi_T(x,\theta),$
where $g$ is a link function,
$\theta = (\beta_0, \beta_1, \beta_2)$ and  $\beta_1, \beta_2 > 0$.  This model forces $\pi_E(x,\theta)$
to be very nonmonotone in $x$.  A more flexible four-parameter model (Thall and Cook, \citeyear{2004Thall}, Section~3)
is given by $\operatorname{Pr}(E|T^c,x,\theta) = g^{-1}(\beta_{E,0}+\beta_{E,1}x)$ and
$\pi_T(x,\theta) = g^{-1}(\beta_{T,0}+\beta_{T,1}x)$,
where $\theta = (\beta_{E,0},\beta_{E,1},\break\beta_{T,0},\beta_{T,1})$
and $\beta_{E,1},\beta_{T,1}>0$.  Using this model,\break  $\pi_E(x,\theta) = \operatorname{Pr}(E|T^c,x,\theta)\{1-\pi_T(x,\theta)\}$.
For either model, $f(Y \vert x, \theta) = \prod_{y=E,N,T}\{\pi_{y}(x,\theta)\}^{I(Y=y)}$.

For the bivariate binary case, the model must specify the four elementary outcome probabilities
$\pi_{a,b}(x,\break \theta) = \operatorname{Pr}(Y_E=a, Y_T=b\vert x,\theta)$
for $a,b\in \{0,1\}$. The\break p.m.f. of a patient treated at dose $x$ is
$f(Y \vert x, \theta) = \prod_{a=0}^1\prod_{b=0}^1
\{\pi_{a,b}(x,\theta)\}^{I(Y_E=a, Y_T=b)}.$
The general approach used by Thall and Cook (\citeyear{2004Thall}) and Thall, Nguyen and Estey (\citeyear{2008Thall})
is to first specify the two marginal dose-outcome distributions
$\pi_k(x,\theta) =\break  g^{-1}\{\eta_k(x,\theta)\}$, in terms of link function $g$
and linear terms $\eta_k(x,\theta)$ for $k = E$ and $T$, and then define
the joint distribution in terms of the marginals.   Temporarily suppressing $x$ and $\theta$,
$\pi_{a,b}$  is determined by  $(a,b,\pi_E, \pi_T, \psi)$, where $\psi$ is an association parameter.
This may be done tractably using a Gumbel distribution,
\begin{eqnarray}\label{bivbin}
\pi_{a,b} &=& \pi_E^a (1-\pi_E)^{1-a}\pi_T^{b}(1-\pi_T)^{1-b}\nonumber
\\
&&{}+ (-1)^{ a+b} \pi_E (1-\pi_E)
\\
&&{}\hspace*{10pt}\cdot \pi_T (1-\pi_T) \biggl(\frac{e^{\psi}-1}{e^{\psi}+1} \biggr),\nonumber
\end{eqnarray}
with $\psi$ real-valued, or a Gaussian copula,
$C(u,v) = \Phi_{\psi}(\Phi^{-1}(u),\Phi^{-1}(v))$, for $0 \leq u,v \leq 1$,
where $\Phi_{\psi}$ is the bivariate standard normal c.d.f. with correlation $\psi$
and $\Phi$ is the univariate $N(0,1)$ c.d.f. Under this copula,
$\pi_{0,0} = \Phi_{\psi}(\Phi^{-1}(1-\pi_E),\Phi^{-1}(1-\pi_T))$
with  $\pi_{1,0} =  1-\pi_T - \pi_{0,0}$,
and $\pi_{1,1} = \pi_E + \pi_T  + \pi_{0,0}- 1$.  If $g$ is the probit link,
$\pi_k = \Phi(\eta_k)$ and $\pi_{0,0} = \Phi_{\psi}(-\eta_E, -\eta_T)$.

A major practical issue is that the $\eta_k(x,\theta)$'s should be realistic but the model must be
numerically trac\-table, to facilitate the processes of fitting historical data if available,
prior elicitation, and computing posterior decision criteria thousands of times
while simulating the trial during the design process.
It often is important to allow $\pi_E(x,\theta)$ to be nonmonotone in $x$,
which may be appropriate for biological agents,
such as viral vectors expressing cytokines aimed at triggering an immune response to kill tumor cells.
This may be done very effectively by assuming a simple quadratic
$\eta_E(x,\theta) = \beta_{E,0} + x \beta_{E,1}  + x^2 \beta_{E,2}$,
although other functions may be used. While $\eta_T(x,\theta) = \beta_{T,0} + x \beta_{T,1}$
with $\operatorname{Pr}(\beta_{T,1}>0) = 1$ is appropriate for cytotoxic agents, in other settings a quadratic also
may be used for $\eta_T(x,\theta)$.  For example, if ``toxicity'' includes
infection and an anti-cancer agent also kills bacterial or fungal infections, then
$\pi_T(x,\theta)$ may be nonmonotone and actually decrease with higher $x$.

Thall and Cook (\citeyear{2004Thall}) provide a penalized least squares method for establishing the prior $p(\theta|\xi)$
based on elicited means
$\mu_{y,j}^{(e)}$ and standard deviations\vspace*{-2pt} (s.d.'s) $\sigma_{y,j}^{(e)}$ of $\pi_y(x_j,\theta)$ for $y = E,T$
and several doses $x_1,\ldots,\break x_m$.
Since each prior mean $\mu_{y,j}(\xi)$ and s.d. $\sigma_{y,j}(\xi)$ of $\pi_y(x_j,\theta)$
is a function of the fixed hyperparameters $\xi$ characterizing $p(\theta|\xi)$,
nonlinear least squares may be used to solve for $\xi$ by minimizing the objective function
\begin{eqnarray} \label{objective}
\operatorname{SS}(\xi) &=& \sum_{y=E,T} \sum_{j=1}^m
   \bigl[ \bigl\{\mu_{y,j}^{(e)}- \mu_{y,j}(\xi)\bigr\}^2\nonumber
   \\
   &&{}\hspace*{46pt}+ \bigl\{\sigma_{y,j}^{(e)}- \sigma_{y,j}(\xi)\bigr\}^2  \bigr]
  \\
  &&{} + c \sum_{1 \leq j < k \leq m} \{\tilde\sigma_{j}-\tilde\sigma_{k}\}^2,\nonumber
\end{eqnarray}
where each $\tilde\sigma_{k}$ is a prior standard deviation in $\xi$.  The second sum in ($\ref{objective}$) is included
to limit the variability among the prior s.d.'s, using a small penalty constant $c>0$.

\subsection{Dose Admissibility and Efficacy-Toxicity
Trade-offs}\label{sec3.2}

The dose-finding algorithm relies on two different types of posterior decision criteria
to choose $x$ from a finite set of possibilities, $\mathcal{X} = \{x^{(1)},\ldots,x^{(k)}\}$.
The first criterion determines which doses are acceptable,
and the second chooses the best acceptable dose.
Let $\underline{\pi}_E$ be a fixed lower limit on ${\pi}_E(x,\theta)$
and $\overline{\pi}_T$ a fixed upper limit  on ${\pi}_T(x,\theta)$.
The fixed limits are specified by the physician.
Let $p_E^*$ and $p_T^*$ both be fixed upper probability cut-offs, usually selected from the range 0.80 to 0.95.
A dose {\it $x$ is unacceptable} if it is likely to have either unacceptably low efficacy or
unacceptably high toxicity, formally if
\begin{eqnarray}\label{stopET}
\operatorname{Pr}\{\pi_E(x,\theta) < \underline{\pi}_E\vert \mathcal{D}_n\} &>& p_E^*  \quad  \mbox{or}\nonumber
\\[-8pt]\\[-8pt]
\operatorname{Pr}\{\pi_T(x,\theta) > \overline{\pi}_T\vert \mathcal{D}_n\} &>& p_T^*.\nonumber
\end{eqnarray}
A  dose $x$ is {\it acceptable} if neither inequality in ($\ref{stopET}$) holds.
The set of acceptable doses in $\mathcal{X}$ based on $\mathcal{D}_n$ is denoted by $\mathcal{A}_n.$
These criteria are essentially those used by Thall, Simon and Estey (\citeyear{1995Thall}) as stopping rules
in phase II trials, and the second criterion in (\ref{stopET}) is used routinely for deciding whether to
stop a phase I trial, for example, when using  the CRM, if the lowest dose is too toxic.
For example, when using EffTox, if $p_E^* = p_T^* = 0.90$,
the rules in ($\ref{stopET}$) are equivalent to saying that $x$ is {\it acceptable} if
$\operatorname{Pr}\{\pi_E(x,\theta) > \underline{\pi}_E\vert \mathcal{D}_n\} > 0.10$
and $\operatorname{Pr}\{\pi_T(x,\theta) < \overline{\pi}_T\vert \mathcal{D}_n\} > 0.10$.  While, intuitively, these may seem
like rather weak requirements, if one replaces 0.10 by a large cut-off such as 0.80 by setting
$p_E^* = p_T^* = 0.20$ in ($\ref{stopET}$), then the rules are nearly certain to stop
any trial very quickly after a very small number of patients,
due to the large variability of the posterior probabilities used in ($\ref{stopET}$).
This gets at the important distinction between determining the acceptability of $x$
for the purpose of dose-finding with small to moderate sample sizes,  and the confirmatory statement
``$x$ is safe and effective'' formalized by inequalities such as
$\operatorname{Pr}\{\pi_E(x,\theta) > \underline{\pi}_E\vert \mathcal{D}_n\} > 0.95$
and $\operatorname{Pr}\{\pi_T(x,\theta) < \overline{\pi}_T\vert \mathcal{D}_n\} > 0.95$.
Such confirmatory conclusions are inappropriate based on early phase trial results
since they can only be established convincingly by a large sample size,
regardless of what the posterior probabilities may be.

\begin{figure*}[t]
\begin{tabular}{cc}

\includegraphics{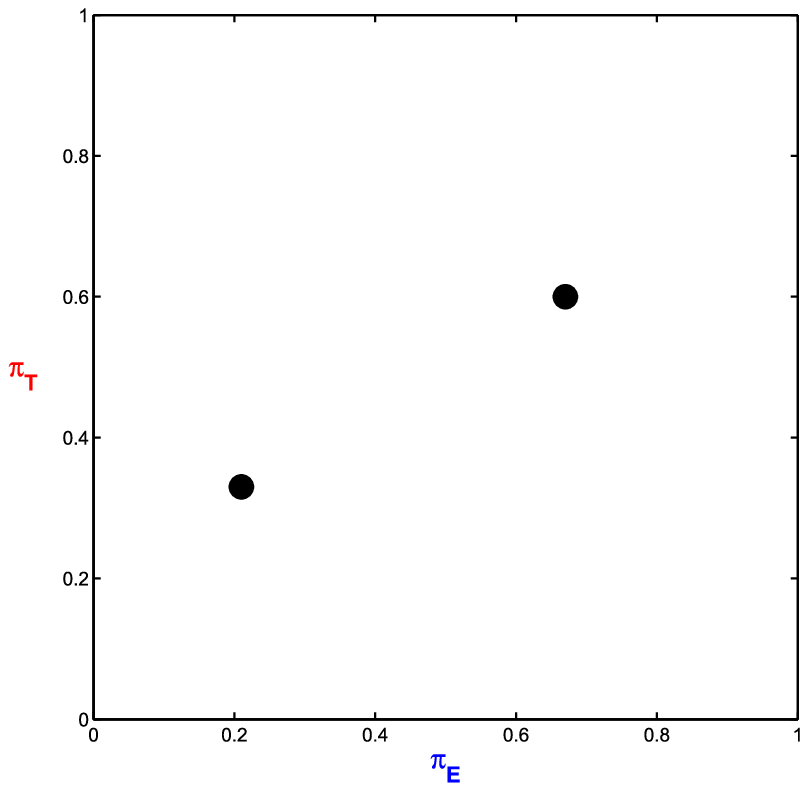}
&\includegraphics{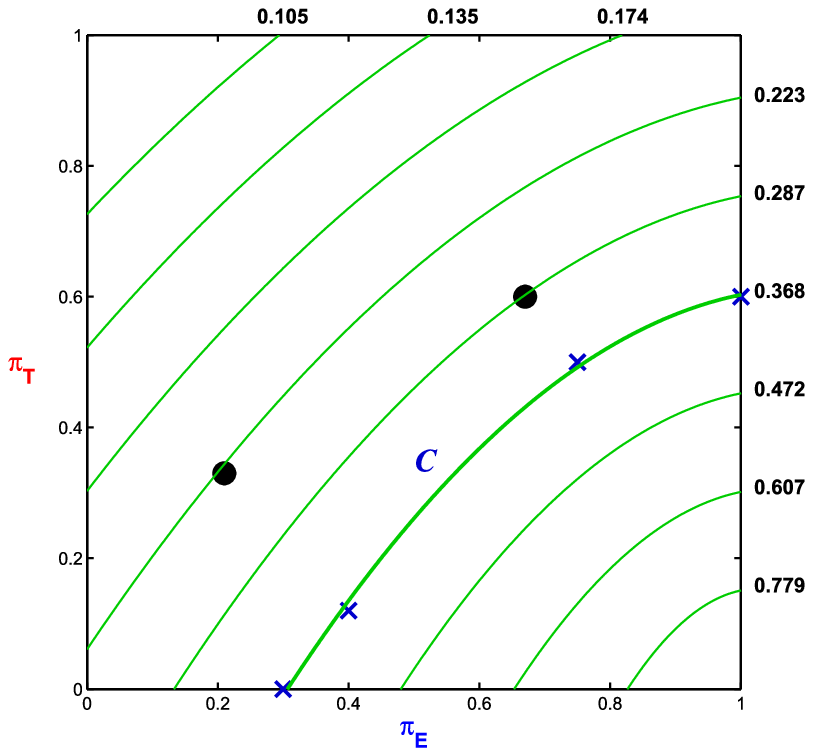}\\
  (a)&(b)
  \end{tabular}
  \caption{Example of posterior means $\mu^{(n)}(x^{(j)}) =
(\mathrm{E}\{\pi_E(x^{(j)},\theta)|\mathit{data}_n\}$, $\mathrm{E}\{\pi_T(x^{(j)},\theta)|\mathit{data}_n\}$)
for two dose pairs $x^{(1)}$ and $x^{(2)}$ (denoted by round dots),
given alone in the left-hand graph (Figure~2a), and
with the addition of  the target contour ${\mathcal C}$ constructed
from elicited target points (denoted by $\times$)
and several resulting desirability contours (Figure~2b).}\label{fig2}
\end{figure*}

To describe the second decision criterion, for simplicity,
I will focus on the bivariate binary case, where  $\pi(x,\theta)\in [0, 1]^2$.
To compare two acceptable doses, say, $x^{(1)}$ and $x^{(2)},$
based on the posteriors of $\pi(x^{(k)},\theta) =  (\pi_E(x^{(k)},\theta),\pi_T(x^{(k)},\theta))$  for $k=1,2$,
some method for reducing each pair $\pi(x^{(k)},\theta)$ to a one-dimensional criterion is required, as
inevitably  is the case when a statistic of dimension $\geq$ 2 is used for comparison.
The EffTox method does this by formalizing the idea that
a higher risk of toxicity is a reasonable trade-off for a  higher probability of
achieving anti-disease effect.
The method first\vspace*{-1pt} computes the posterior means $\mu^{(n)}(x)
= (\mu_E^{(n)}(x),\mu_T^{(n)}(x))  = (\mathrm{E}\{\pi_E(x,\theta)|\mathcal{D}_n\}, \mathrm{E}\{\pi_T(x,\theta)|\mathcal{D}_n\})$
for each $x\in\mathcal{X}$.\break  Each $\mu^{(n)}(x)$ is then reduced to a one-dimensional criterion
by the following geometric construction, which
begins by eliciting several pairs of fixed probabilities, $p^{(j)} = (p_E^{(j)},p_T^{(j)})$,
$j=1,\ldots,m$,
that the physician considers equally desirable.    A target curve, $\mathcal{C}$, is fit
to the elicited pairs, treating $p_T$ as a monotone increasing function of $p_E$,
or, equivalently, reversing the roles of  $p_T$ and  $p_E$.
This should be done using a graphical representation of $\mathcal{C}$ to provide a means for the
physician to adaptively modify his/her target pairs.
Given  $p\in [0, 1]^2$, let $p_\mathcal{C}$ denote the point where
the straight line segment in $[0, 1]^2$ passing through $p$ and the ideal point (1, 0) intersects $\mathcal{C}$.
The  {\it desirability of $p$} may be defined as
\begin{equation}\label{deltadef}
\delta(p) =
\exp \biggl\{- \frac{ \| p - (1,0) \|}{\| p_\mathcal{C} - (1,0) \| } \biggr\},
\end{equation}
where  $\|\cdot \|$ denotes Euclidean distance.
This has maximum $\delta(1,0) = 1$, with $\delta(p)$ decreasing as $p$ moves away
from (1, 0) along any straight line in $[0, 1]^2$. Several other definitions of $\delta(p)$ may be used,
although ($\ref{deltadef}$) is reasonable and tractable. The contour of all $p$ having desirability $u$ is
$\mathcal{C}_u = \{p\in [0, 1]^2\dvtx \delta(p) = u\}$, so that $\mathcal{C} = \mathcal{C}_{e^{-1}}$.
Denote the set of real numbers $u$ such that $C_u \neq \phi$
by $R_\mathcal{C}$. Since $u\neq v \Longrightarrow \mathcal{C}_u\cap \mathcal{C}_v = \phi$, the family
$\{\mathcal{C}_u, u\in R_\mathcal{C}\}$ partitions $[0, 1]^2$.  This construction is used to
quantify the desirability of a dose $x$ by evaluating ($\ref{deltadef}$) at $p=\mu^{(n)}(x)$.
To compare doses $x^{(1)}$ and $x^{(2)}$, we compute  $\mu^{(n)}(x^{(1)})$
and $\mu^{(n)}(x^{(2)})$, illustrated by the two round dots in
Figure~\ref{fig2}a, and then compute their desirabilities $\delta(\mu^{(n)}(x^{(1)}))$
and $\delta(\mu^{(n)}(x^{(2)}))$, as shown in Figure~\ref{fig2}b.  The elicited pairs are
represented by the symbol ``$\times$'' in Figure~\ref{fig2}b, which also shows $\mathcal{C}$
and several $\mathcal{C}_u$ along with their numerical $u$ values.  During the trial,
if no dose is acceptable, formally if $\mathcal{A}_n = \phi$, then accrual is stopped with no dose selected;
otherwise each cohort is given the dose $x$ maximizing $\delta(\mu^{(n)}(x))$ among all
$x\in \mathcal{A}_n$. This methodology has been used for dose-finding trials in acute stroke,
treatment for GVHD in SCT, chemotherapy of acute leukemia,
and anergized cells given post-transplant to accelerate immune reconstitution following
allotx.

It is important to consider the consequences of how one sets goals
in the bivariate binary outcome case. Recall that $\pi_{a,b} = \operatorname{Pr}(Y_E=a,Y_T=b)$
with $\pi_E = \pi_{1,1}+ \pi_{1,0}$ and $\pi_T = \pi_{1,1}+ \pi_{0,1}$.
Several authors have proposed choosing $x$ to maximize $\pi_{1,0},$
the probability of the best possible outcome, efficacy and no toxicity,
or the conditional probability $\pi_{E | T^c}  = \pi_{1,0}/(1-\pi_T)$.
Unfortunately, most new treatments simply don't work that way.
A new therapy that is either more aggressive, for example, a higher dose, or highly active biologically
is likely to decrease $\pi_{0,0}$ and increase some combination of  $\pi_{1,0}, \pi_{0,1}$
and $\pi_{1,1}$. Treating $\pi = (\pi_{0,0},\pi_{1,0},\pi_{0,1},\pi_{1,1})$ as fixed
for simplicity, suppose that standard treatment gives outcome probability vector
$\pi^{(0)} = (0.50, 0.10, 0.30,0.10)$, which has
marginals $(\pi_E, \pi_T) = (0.20, 0.40)$.  Suppose that experimental treatment $x^{(1)}$
has $\pi^{(1)} =  \pi(x^{(1)}) = (0.30, 0.20, 0.30, 0.20)$, which has
marginals $(\pi_E^{(1)},\break\pi_T^{(1)}) = (0.40, 0.50)$,  a doubling of $\pi_E^{(0)}$ from 0.20 to 0.40
and a 25\% increase in $\pi_T^{(0)}$ from 0.40 to 0.50.   Suppose that a competing experimental treatment $x^{(2)}$
has $\pi^{(2)} =  \pi(x^{(2)}) = (0.30, 0.20, 0.45, 0.05$), which has
marginals $(\pi_E^{(2)},\pi_T^{(2)}) = (0.25, 0.50)$,  a slight increase in $\pi_E^{(0)}$ from 0.20 to 0.25
with the same increase in $\pi_T^{(0)}$ as given by $x^{(1)}$.
Since $\pi_{1,0}(x^{(1)}) = \pi_{1,0}(x^{(2)}) =  0.20$ and
$\pi_{E | T^c}(x^{(1)}) = \pi_{E | T^c}(x^{(2)}) = 0.40$, a method based on
either $\pi_{1,0}$  or $\pi_{E | T^c}$ would consider $x^{(1)}$  and $x^{(2)}$  to be equivalent.
In contrast, the trade-off based method  would consider $x^{(1)}$ superior to $x^{(2)}$.

\section{Finding Patient-Specific Doses}

Thall, Nguyen and Estey (\citeyear{2008Thall}) generalized EffTox
to account for patient heterogeneity by using the patient's vector
$Z = (Z_1,\ldots,Z_q)$ of  covariates observed at enrollment.
The method requires historical data, $\mathcal{H}$, to obtain an informative distribution on
covariate effect parameters for use in trial design and conduct.  The model and method account for
dose effects, covariate effects and possible dose-covariate interactive effects on $\pi_E$ and $\pi_T$.
The design assigns each patient a dose that is individualized based on the patient's $Z$ vector.
This is very different from conventional early phase trial designs, since (i) patients with different
covariates may receive different doses at the same point in the trial, (ii) the entry criteria may change adaptively,
with the possibility that enrollment may be shut down for some patients but continued for others,
and (iii) at the end of the trial a computer-based rule is provided for assigning each future patient's
$x$ based on his/her $Z$ vector, rather than choosing a single dose for all patients.

For designs with individualized treatment assignment rules
utilizing $Z$ (cf. Ratain  et al., \citeyear{1996Ratain}; Babb and Rogatko, \citeyear{2001Babb}),
the $i${th} patient's data are\vspace*{1pt} $(x^{(i)},\break Z^{(i)},Y^{(i)})$,
the probability model is elaborated by defining $f(y|Z,x,\theta)$
for a patient with covariates $Z$ who receives treatment $x$,
and $\alpha_n$ is a function of $(Z,\mathcal{D}_n)$.  To accommodate $Z$ and historical data
in the design described here,
let $\tau$ denote either a dose $x$ in the trial or historical treatment from the set
$\{\tau_1,\ldots,\tau_m\}$.
The probability model given earlier is extended by defining the marginal probabilities,
$\pi_k(\tau,Z,\theta) = g^{-1}\{\eta_k(\tau,Z,\theta)\}$, $k=E,T,$
for a patient with covariates $Z$ given dose $x$,   assuming linear terms of the general form
\begin{eqnarray}\label{linearETZ}
\quad \eta_k(\tau, Z,\theta)&=& \beta_k Z +
\sum_{j=1}^m (\mu_{k,j}+ \xi_{k,j} Z ) I(\tau=\tau_j)\nonumber
\\[-8pt]\\[-8pt]
&&{}+
\{\omega_k(x, \alpha_k) + \gamma_k Z \}I(\tau=x)\nonumber
\end{eqnarray}
for $k = E,T$, where
$\beta_k Z = \beta_{k,1}Z_1+\cdots+\beta_{k,q}Z_q$ account for covariate main effects,
$\gamma_k Z = \gamma_{k,1}Z_1+\cdots+\gamma_{k,q}Z_q$ account for  dose-covariate interactions,
$\mu_k = (\mu_{k,1},\ldots,\mu_{k,m})$ are historical main treatment effects,
$\xi_{k,j} Z = \xi_{k,j,1}Z_1+\cdots+\xi_{k,j,q}Z_q$ account for covariate interactions with the $j${th} historical
treatment, and $\omega_E(x, \alpha_E)$ and $\omega_T(x, \alpha_T)$ are the usual dose-outcome functions characterizing
main dose effects, and  $\omega_E$ and $\omega_T$ may be
quadratic or linear functions of $x$, as given earlier.
For fitting the historical data, ($\ref{linearETZ}$) takes the form
\begin{eqnarray}
&&\eta_k(\tau_j, Z,   \theta) = \mu_{k,j} + \beta_{k}  Z +  \xi_{k,j} Z\nonumber
      \\[-8pt]\\[-8pt]
      &&\quad \mbox{for } j=1,\ldots,m   \mbox{ and } k = E,T.\nonumber
\end{eqnarray}
For fitting the data obtained during the trial, ($\ref{linearETZ}$) is
\begin{eqnarray}\label{linearT}
&&\eta_k(x, Z, \theta) =
\omega_k(x, \alpha_k) + \beta_k  Z + x  \gamma_k  Z\nonumber
\\[-8pt]\\[-8pt]
&&\quad    \mbox{for } k = E,T.\nonumber
\end{eqnarray}

A much more parsimonious model that accounts for dose-covariate interactions is obtained by replacing
$x  \gamma_k  Z$ in ($\ref{linearT}$) with either
$\gamma_k\{\omega_k(x, \alpha_k)\times \beta_k  Z\}$ or $\gamma_k\{x\times \beta_k  Z\}$
where each $\gamma_k$ is now a single parameter rather than a $q$-dimensional vector.   This
model requires only 2 dose-covariate interaction parameters instead of $2q.$
This is motivated by the idea that $\gamma_k  \{\omega_k(x, \alpha_k)\times \beta_k  Z\}$
is similar to the one-degree of freedom interaction term in the model for a two-way layout with one observation per cell
given by Tukey (\citeyear{1949Tukey}).  Unfortunately, in practice,
this parsimonious model is a complete disaster since, using either $\omega_k(x, \alpha_k)$ or $x$,
it gives a very poor fit to the trial data
when dose-covariate interactions of any complexity are present.  So this more parsimonious model
is a cute idea that simply doesn't work.

Generalizing the EffTox design to accommodate $Z$ requires much more than writing down a model.
The set $\mathcal{A}_n(Z)$ of acceptable doses
for a patient with covariates $Z$ is defined to be all $x\in\mathcal{X}$ satisfying the constraints
\begin{eqnarray}\label{ADET}
\qquad \operatorname{Pr}\{\pi_E(x,Z,\theta) < \underline{\pi}_E(Z)\vert \mathcal{D}_n\cup\mathcal{H}\} &<& p_E^*\quad\mbox{and}\nonumber
\\[-8pt]\\[-8pt]
\operatorname{Pr}\{\pi_T(x, Z, \theta) > \overline{\pi}_T(Z)\vert \mathcal{D}_n\cup\mathcal{H}\} &<& p_T^* ,\nonumber
\end{eqnarray}
where $\underline{\pi}_E(Z)$ and $\overline{\pi}_T(Z)$ are {\it acceptability bounding functions},
constructed as follows.  First, a
representative set of covariate vectors, $\{ Z^{(1)},\ldots, Z^{(K)}\},$
is determined.
For each $Z^{(j)}$, the physician specifies the smallest probability of efficacy, $\underline{\pi}_E^{(j)}$,
and the largest probability of toxicity, $\overline{\pi}_T^{(j)}$,
that are acceptable for a patient having those covariates.
For $k=E,T,$ denote $\zeta_k(Z) = \mathrm{E}(\beta_k Z\vert \mathcal{H})$,
the historical posterior mean of the covariate main effect linear combination.
To construct the bounding function $\underline{\pi}_E(Z)$ for $\pi_E(x, Z, \theta)$, the $K$
pairs
$(\zeta_E( Z^{(1)}),\underline{\pi}_{E}^{(1)}),\ldots,\break
(\zeta_E( Z^{(K)}),\underline{\pi}_{E}^{(K)})$
of estimated linear terms and elicited lower bounds on $\pi_E$ are used as
regression data to fit a simple linear or quadratic curve by least squares,
using $\zeta_E( Z^{(j)})$ as the predictor and $\underline{\pi}_E^{(j)}$ as the outcome variable.
Denoting the estimated outcome under the fitted regression model by $\hat{\underline{\pi}}_E({\zeta}_{E}(Z))$, the
{\it efficacy lower bounding function}  is
${\underline{\pi}}_E( Z) = \hat{\underline{\pi}}_E\circ\zeta_E( Z)$.
The {\it toxicity upper bounding function} ${\overline{\pi}}_T( Z) = \hat{\overline{\pi}}_T\circ\zeta_T( Z)$
is computed similarly  from $(\zeta_T( Z^{(1)}),\break\overline{\pi}_{T}^{(1)}),\ldots,
(\zeta_T(Z^{(K)}),\overline{\pi}_{E}^{(K)})$.
When constructing these functions, it is important to plot the scattergrams of
the constructed regression data sets along with the fitted curves, which
the physician may use to guide adjustment of some $\underline{\pi}_E^{(j)}$ or $\overline{\pi}_T^{(j)}$ values,
if desired, to obtain acceptability bounding functions
${\underline{\pi}}_E( Z)$ and ${\overline{\pi}}_T( Z)$
that make sense clinically.  These constructions
map each patient's $Z$ vector into the probability bounds used in ($\ref{ADET}$) to determine
whether each $x\in\mathcal{X}$ is acceptable for that patient.
To define a covariate-specific dose desirability index, we evaluate $\delta(p)$ given by ($\ref{deltadef}$)
at $p = \mu^{(n)}(x,Z)  = (\mathrm{E}\{\pi_E(x,Z,\break\theta)|\mathcal{D}_n\}, \mathrm{E}\{\pi_T(x,Z,\theta)|\mathcal{D}_n\}),$
and denote this by\break $\delta_n(x,Z)$.
For two patients with different covariates $Z_1 \neq Z_2$, it may be the case that $\mathcal{A}_n(Z_1)\neq \mathcal{A}_n(Z_2)$,
including the possibility that $\mathcal{A}_n(Z)=\phi$ for one patient but not the other.
Even if $\mathcal{A}_n(Z_1) = \mathcal{A}_n(Z_2)$,
the $x$ that maximizes $\delta_n(x,Z_1)$ may not be the same as that maximizing
$\delta_n(x,Z_2)$.

\begin{figure*}[t]

\includegraphics{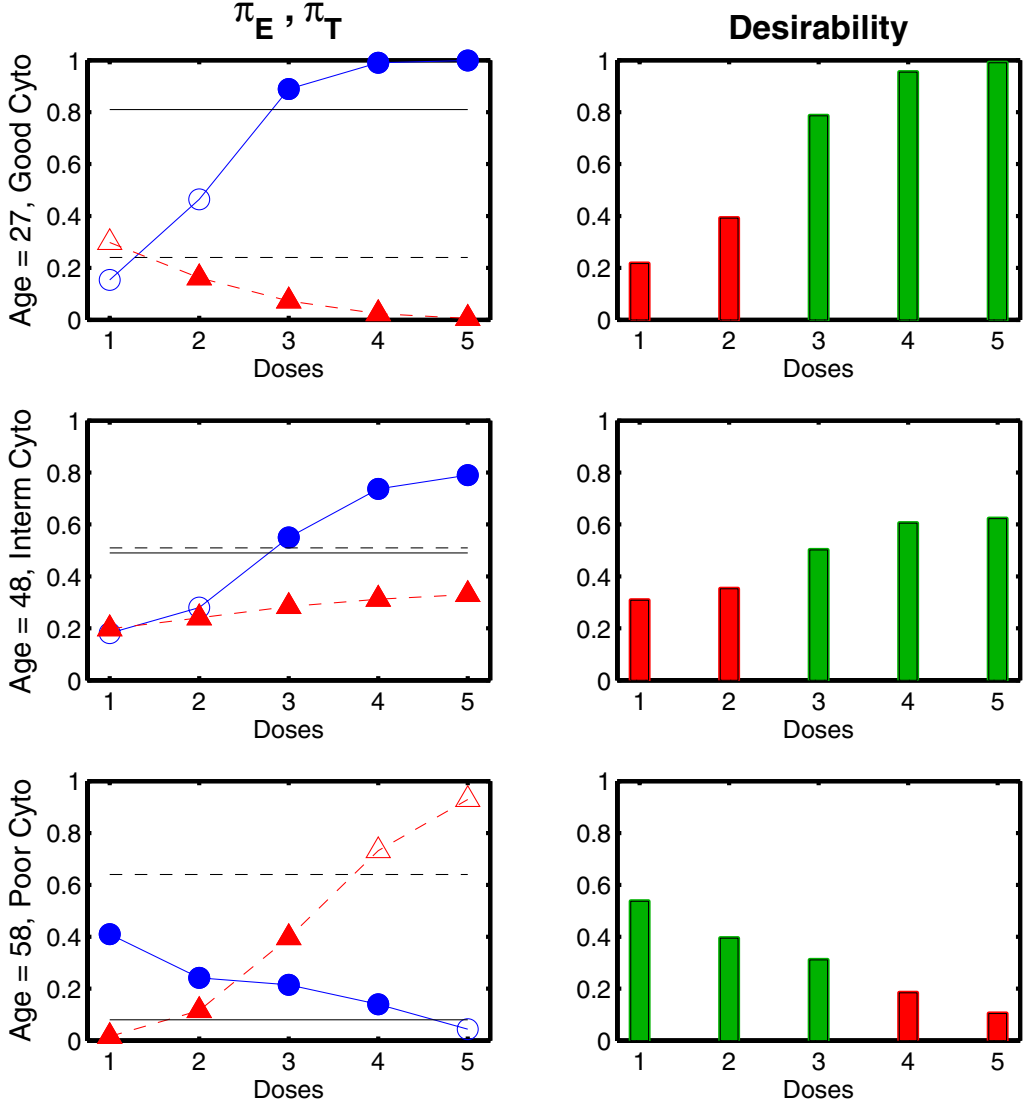}

  \caption{Marginal outcome probabilities $\pi_E(x,\theta)$, given as circles,
and $\pi_T(x,\theta)$, given as triangles (left column) and corresponding desirabilities (right column)
for each of three different patient prognostic vectors,
under a fixed $\theta$. Solid (open) points correspond to acceptable (unacceptable) doses.}\label{fig3}
\end{figure*}

Figure~\ref{fig3} illustrates how the dose-efficacy and dose-toxicity probability curves in $x$ also
may change with~$Z$.
The curves are given for a particular fixed $\theta^{\mathit{true}}$ in which the interactive effects of $x$ and $Z$ are substantial,
taken from the acute leukemia application discussed by Thall, Nguyen and Estey (\citeyear{2008Thall}) where $Z = {}$(AGE, cytogenetic abnormality),
with the second covariate coded as a three-category variable having possible
values \{Good, Intermediate, Poor\} defined in terms of prognostic level.
The rows in Figure~\ref{fig3} correspond to three different $Z$ values. In the left column,
the probabilities $\pi_E(x,\theta)$ are represented by circles and
$\pi_T(x,\theta)$ by triangles,
with an open (filled) circle or triangle representing an unacceptable (acceptable) dose.
The corresponding desirabilities are given in the right column,
obtained by evaluating $\delta(\pi_E(x,Z,\theta^{\mathit{true}}),\pi_T(x,Z,\theta^{\mathit{true}})).$
The figure shows that the dose-outcome functions $\pi_E(x,\theta)$ and $\pi_T(x,\theta)$ may change dramatically with $Z$,
that the effect of prognosis may be as large as or larger than that of dose,
and that interactions between $x$ and $Z$ may be quite important.  Figure~\ref{fig3} also illustrates
how the desirability function $\delta$ reduces each two-dimensional $(p_E,p_T)$ to a one-dimensional value
that may be used to compare doses for each $Z$.

To apply this methodology, the first step is to analyze~$\mathcal{H}$ under
several models, choose the model providing the best fit, compute
$p(\bolds{\beta},\psi \vert \mathcal{H})$, and determine noninformative priors on
$\bolds{\alpha}$ and $\bolds{\gamma}$.  During the trial, when a patient with covariates
$Z$ is enrolled, $\mathcal{A}_n(Z)$ is computed. If $\mathcal{A}_n(\mathbf{ Z}) =\phi$, the patient is not treated on protocol.
If $\mathcal{A}_n(\mathbf{ Z})\neq \phi$, the patient is
treated with the dose $x$ maximizing $\delta_n(x,Z).$  If
$\mathcal{A}_n(Z^{(j)}) = \phi$ for all representative covariates,
then the trial is stopped. After the trial, given final data $\mathcal{D}_N$, the
decision rules based on
$p(\bolds{\theta} \vert \mathcal{H}\cup\mathcal{D}_N) $ are used to select doses for future patients.

Our computer simulation studies of this new\break methodology have produced some disquieting messages.
The first is that ignoring established prognostic covariates may lead to either very
unsafe or very ineffective dose assignments for many patients both during and after a phase I or phase I/II trial.
The second message is that, if dose-covariate interactions are
present, ignoring them by using an additive model for the effects of $x$ and $Z$ also may lead to very
poor dose assignments.  That is, the common practice of ignoring known patient heterogeneity in early
phase trials may lead to bad science and bad clinical practice.

\section{Accounting for Multiple Toxicities}\label{sec5}

\subsection{Outcomes and Model}

Bekele and Thall (BT, \citeyear{2004Bekele}) proposed a dose-finding method
based on a vector $Y = (Y_1,\ldots,Y_J)$ of several qualitatively different types of toxicity,
with $Y_j$ an ordinal variable recording the $j${th} toxicity's severity. The method
was motivated by a phase I trial to choose a dose of  gemcitabine, in mg$/$m$^2$,
from $\{100, \ldots,  1000\}$
when combined with a fixed dose of 50 cGy external beam  radiation, both
given prior to surgery, for patients with soft tissue
sarcoma.  The design was developed working with a team of three oncologists
who had extensive experience treating sarcomas.  The  point of departure
from conventional methods is that the design distinguishes between different types of toxicity,
and it also accounts for the severity levels of each.

Denote the $m_j+1$ severity levels of $Y_j$ by $\{y_{j,0}, y_{j,1},\break\ldots, y_{j,m_j}\}$.
For example, in the sarcoma trial the 4 levels of liver toxicity were $y_{j,0} = {}$\{grade 0 or~1\},
$y_{j,1} = {}$\{grade 2\}, $y_{j,2} = {}$\{grade 3\} and $y_{j,3} = {}$\{grade~4\}.
Binary $Y_j$  corresponds to $m_j = 1$.  Using standardized doses
$x = \log$\{(raw dose)$/$1000\}, so that $\mathcal{X} = \{-2.30, -1.61, \ldots, 0\}$,
the distribution of $Y|x$
was modeled using the method of Albert and Chib (\citeyear{1993Albert}),
in terms of the $J$-vector of Gaussian latent variables
$\zeta = (\zeta_1,\ldots,\zeta_J)$ with $\mathrm{E}(\zeta_j|x) = \beta_{j,0}+x\beta_{j,1}$,
$\operatorname{var}(\zeta_j) = 1$ and correlation matrix $\Omega$, by
defining
$Y_j = y_{j,k}$ if $\gamma_{j,k} \leq \zeta_j < \gamma_{j,k+1}$
for $k = 0, 1, \ldots, m_j$ and $j = 1,\ldots,J$
for cut-off parameters $\gamma_{j} = (\gamma_{j,1},\ldots,\gamma_{j,m_j})$
satisfying $-\infty = \gamma_{j,0} < \gamma_{j,1} <\cdots < \gamma_{j,m_j} <\gamma_{j,m_{j}+1} = +\infty,$
with $\gamma_{j,1} \equiv 0$ to ensure identifiability.
This formulation greatly facilitates MCMC computations used to obtain posterior quantities.
Denoting the $2J$-vector of regression parameters
$\beta = (\beta_{1,0},\beta_{1,1},\ldots,\beta_{J,0}, \beta_{J,1}),$
 the vector  $\gamma = (\gamma_1,\ldots,\gamma_J)$
having $m_+ = m_1+\cdots + m_j$ entries, and
the $J(J-1)/2$ off-diagonal elements of
$\Omega$ by $\rho = (\rho_{1,2},\rho_{1,3},\ldots,\rho_{J-1,J})$,
the model parameter vector is $\theta = (\beta,\gamma,\rho).$
The marginal distribution of $Y_j|x$ is  given by
\begin{eqnarray}
\pi_{j,k}(x,\theta) &=& \operatorname{Pr}(Y_j = y_{j,k} \vert x,\theta)\nonumber
\\
&=& \Phi(\gamma_{j,k+1} -\beta_{j,0}-\beta_{j,1}x)
\\
&&{}- \Phi(\gamma_{j,k} -\beta_{j,0}-\beta_{j,1}x).\nonumber
\end{eqnarray}

To obtain an expression for the joint distribution,
denote the p.d.f. of a multivariate normal random vector $W$
with mean vector $\mu$ and variance--covariance matrix $\Sigma$
by $\phi_W(\cdot \vert \mu,\Sigma)$.  In matrix notation,
$\mathrm{E}(\zeta|x) = X \beta'$,  where $X$ is the $J \times 2J$
block diagonal matrix with $J$ identical blocks $(\matrix{1 & x})$.
Denote the  intervals  $G_{j,k} = (\gamma_{j,k}, \gamma_{j,k+1}]$.
For observed vector $k = (k_1,\ldots,k_J)$
of toxicity severity levels, the outcome is $Y = y(k) =
(y_{1,k_1},\ldots,y_{J,k_J}),$ which corresponds to latent $\zeta$ values
in the $J$-dimensional set $G(k,\gamma) =
G_{1,k_1}\times\cdots\times G_{J,k_J}$.
A single patient's likelihood contribution is
\begin{eqnarray}
 \quad &&\mathcal{L}(Y\vert x, \theta)\nonumber
\\
 &&\quad= \prod_{k_1=0}^{m_1}\cdots
 \prod_{k_J=0}^{m_J}
\biggl\{ \int_{ G( k, \gamma)} \phi_Z( z\vert  X \beta',
 \\
 &&\qquad\hspace*{109pt} {}\Omega )\, d z \biggr\}^{I[Y=y(k)]}.\nonumber
\end{eqnarray}

For priors, BT assume $\beta \sim N(\mu,\Sigma)$,
subject to $\operatorname{Pr}(\beta_{j,1} > 0) = 1$ for all
$j=1,\ldots,J,$ so that  $\beta$ is $2J$-variate normal with
all slope coefficients truncated below at
0, but $\mu$ and $\Sigma$ correspond to  the untruncated $2J$-variate
normal.   This ensures that $\operatorname{Pr}(Y_j > y_{j,k} \vert x, \beta) =
1- \Phi\{\gamma_{j,k} - \beta_{j,0} - \beta_{j,1}x\}$ increases with $x$
for each $j$ and  $k>1$.   For each $j$ with $m_j \geq 2$ (3 or more levels),
$\{\gamma_{j,2},\ldots,\gamma_{j,m_j}\}$ follow independent, uninformative priors
on the domain $[0, 10],$ with each $p(\gamma_{j,k}) \propto 1,$ subject to the constraint
$0 < \gamma_{j,2} < \gamma_{j,3} < \cdots  < \gamma_{j,C_j},$
where the upper limit 10 on the support of each
$p(\gamma_{j,k})$ was chosen for numerical convenience.
The $\rho_{j,k}$'s are assumed to be i.i.d. $N(0,1000),$ truncated to have support $[-1, +1],$
with $\Omega$ positive definite.

\begin{figure*}[t]

\includegraphics{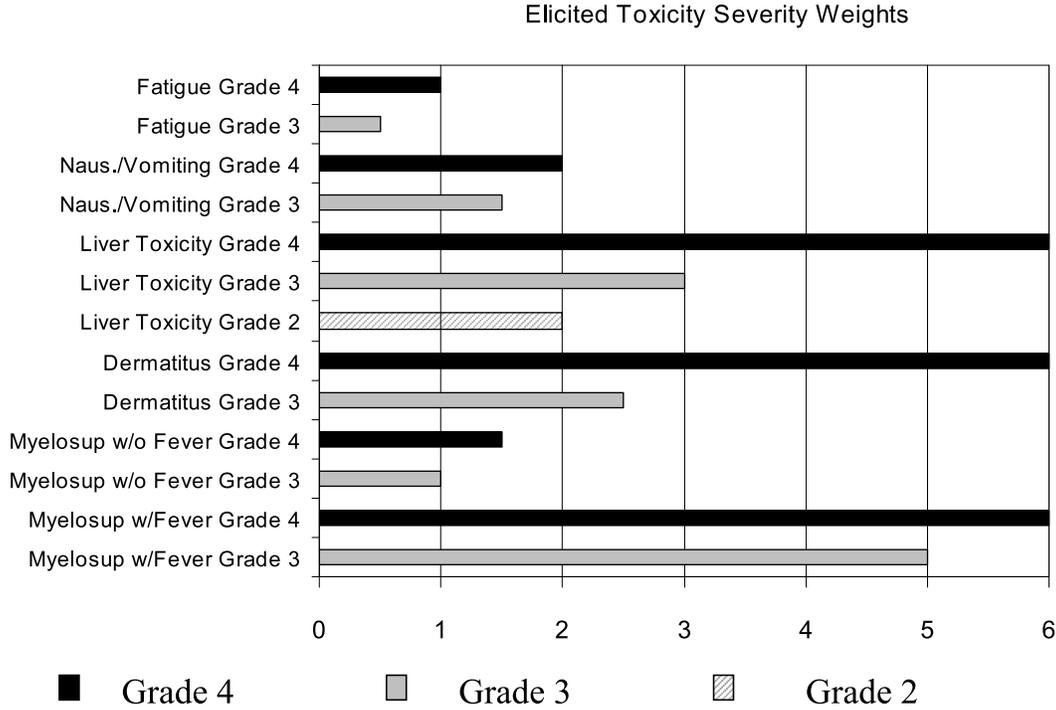}

  \caption{The elicited toxicity severity weights used in the soft tissue sarcoma trial.}\label{fig4}
\end{figure*}

\subsection{Total Toxicity Burden and Trial Conduct}

The dose-finding method is based on {\it toxicity severity weights},  elicited
as follows. The oncologists are first asked to  specify the $J$ toxicities
to be monitored, including the severity levels of each. They are then asked to
specify a numerical severity weight for each level
of each toxicity within a positive-valued numerical range
with which they are comfortable, such as 0 to 10, or 0 to 100.
The severity weights are denoted by $w
= (w_1,\ldots,w_J),$ where $w_j = (w_{j,0},\break w_{j,1},\ldots, w_{j,m_j})$
are the severity weights of the possible values $(y_{j,0}, y_{j,1},\ldots, y_{j,m_j})$ of $Y_j,$
with the obvious requirement $w_{j,0} < w_{j,1} < \cdots  < w_{j,m_j};$
otherwise, if $w_{j,k} = w_{j,k+1}$, then levels $k$ and $k+1$ of $Y_j$ should be combined.
The elicited severity weights used in the sarcoma trial are illustrated
in Figure~\ref{fig4}.  An interesting practical point arose while assigning
severity weights to myelosuppression, which is
defined in terms of low blood cell counts and is caused by effects of chemotherapy on the bone marrow.
At first, no distinction was made between
myelosuppression
occurring either with or without fever.  During the process of establishing $w$, however, the oncologists explained
that myelosuppression is much more severe when it occurs with fever, since it is then
life-threatening and may be an impediment to further chemotherapy.   This led us to redefine
myelosuppression occurring without or with fever as two different types of toxicity.
Figure~\ref{fig4} shows the severity weights 1 and 1.5 for grade 3 and 4 myelosuppression without fever, compared to
weights 5 and 6 for grade 3 and 4 myelosuppression with fever.  Thus, in general,
$Y$ and $w$ are elicited together, and this process is not unlikely to involve iteration.

A patient's {\it total toxicity burden} (TTB) is defined to be
\begin{equation}
\mathit{TTB} = \sum_{j=1}^J\sum_{k=1}^{m_j} w_{j,k} I(Y_j=y_{j,k}).
\end{equation}
For example, from Figure \ref{fig4}, a patient with grade 3 fatigue,
grade 3 nausea/vomiting and grade 4 myelosuppression without fever would have
$\mathit{TTB} = 0.5 + 1.5 + 1.5 =  3.5$, whereas a patient with grade 4 myelosuppression
with fever would have $\mathit{TTB} = 6.0$. Using the
conventional approach of defining a single binary
outcome  $Y$ indicating  at least one grade 3 or 4 toxicity, these two patients
would be scored identically, with both having $Y = 1$.

The posterior expected TTB of dose $x$ is
\begin{eqnarray}
\tau(x,\mathcal{D}_n) &=& \mathrm{E} \bigl(\mathit{TTB} \vert  x,\mathcal{D}_n\bigr)\nonumber
\\[-8pt]\\[-8pt]
&=& \sum_{j=1}^J\sum_{k=1}^{m_j} w_{j,k} \mathrm{E}\{\pi_{j,k}(x,\theta)\vert \mathcal{D}_n\}.\nonumber
\end{eqnarray}
The  trial is conducted by establishing a targeted total toxicity burden, $\mathit{TTB}^*$, and choosing
each cohort's $x$ to minimize $|\tau(x,\mathcal{D}_n) - \mathit{TTB}^*|$.
This is analogous to choosing a dose, based on a binary $Y$ with $\pi(x,\theta) = \operatorname{Pr}(Y=1|x,\theta)$,
using the CRM criterion to minimize $|\mathrm{E}\{\pi(x,\theta)|\mathcal{D}_n\} - \pi^*|$ for given target probability $\pi^*$.
It is easy to show that, since  $w_{j,k-1} < w_{j,k}$ and $\beta_{j,1}>0$
for all $j$ and $k,$  $\tau(x,\mathcal{D}_n)$ is increasing in $x$, so
$x$ may be determined by a monotone search.
The process proposed by BT for establishing the target $\mathit{TTB}^*$ is straightforward, albeit somewhat elaborate.
The physicians are first asked to specify a set of hypothetical patient cohorts
and toxicity outcomes for each patient in each cohort, with the cohorts defined so that the toxicity severities
vary substantially between cohorts.  BT provide a detailed description of this process,
and in the sarcoma trial there were 16 hypothetical cohorts of 4 patients each
with the mean TTB of each cohort varying from $\overline{\mathit{TTB}} = 1.25$ to 5.62.  For each hypothetical cohort,
the oncologists are asked whether observing its toxicity outcomes would lead them
to escalate, repeat the current dose, or de-escalate for the next cohort. The target $\mathit{TTB}^*$
is then defined as the mean of the $\overline{\mathit{TTB}}$ values for which the decision would be to repeat the current dose.
For the sarcoma trial, this yielded  $\mathit{TTB}^* = 3.04$.
Computer simulations of this methodology provided by BT show that it has remarkably attractive OCs
and makes decisions very differently from conventional phase I designs.  A cohort of four  patients all with
myelosuppression grade 4 without fever, patients \#1, \# 2 and \# 3 with grade 3 fatigue,
 and patient \#4 with grade~3
nausea/vomiting would have $\overline{\mathit{TTB}}  = \{(1.5 + 0.5) + (1.5 + 0.5) + (1.5 + 0.5) + (1.5 + 1.5)\}/4
= 2.25$.  The three oncologists all agreed that the appropriate decision based on these outcomes
would be to escalate, whereas any conventional method based on one binary toxicity indicator
would score this as 4 ``toxicities'' in 4 patients and certainly would
\mbox{de-escalate}.

\section{Optimizing Dose and Schedule}

\subsection{A New Paradigm for Phase I Trials}

Braun et al. (BTND, \citeyear{2007Braun}) proposed a new\break paradigm for phase I trials that
jointly optimizes schedule of administration and per-administration dose (PAD) based
on time-to-toxicity. This extends Braun, Yuan and Thall (\citeyear{2005Braun}),
who optimized schedule while assuming a fixed PAD.
Although the model used by BTND is very different from that underlying the TiTE CRM
(Cheung and Chappell, \citeyear{2000Cheung}) for dose-finding based on time-to-toxicity,
the BTND method is a practical extension in that it allows schedule as well dose to be varied.
The treatment regime is
$x = (s, d_s)$, where  $s = (s_1,\ldots,s_k)$ are successive administration  times and $d_s
= (d({s_1}),\ldots,d({s_k}))$ are the doses given at those times.
BTND address the problem of evaluating a $K\times J$ matrix of  $K$ nested
schedules, $s^{(1)}\subset s^{(2)}\subset \cdots \subset s^{(K)},$  where the $k$th schedule is
$s^{(k)} = (s_1, s_2, \ldots, s_{m^{(k)}}),$  so that $m^{(1)} < m^{(2)}< \cdots < m^{(K)}$,
and $J$ PADs, $d^{(1)} < d^{(2)} < \cdots < d^{(J)}$.  The treatment set evaluated by the design is
$\mathcal{X} = \{(s^{(k)},d^{(j)})\dvtx  k=1,\ldots,K, j=1,\ldots,J\}$, and
the total amount of the agent given to the patient increases with both dose and schedule.
For example, a patient assigned PAD $d^{(3)}$ under schedule $s^{(2)} = (s_1, s_2, \ldots,
s_{m^{(2)}})$ receives total dose $d^{(3)} m^{(2)}$ of the agent in $m^{(2)}$ successive administrations of $d^{(3)}$ each,
unless therapy is terminated early due to toxicity, so the planned $d_{s^{(2)}}$
in $x = (s^{(2)}, d_{s^{(2)}})$ is the $m^{(2)}$-vector with all entries $d^{(3)}$.

For this regime, it is helpful to distinguish between two time scales, {\it study time} and {\it patient time}.
Starting at study time 0 when the trial begins,
let $e$ be a given patient's entry time, so that the patient's assigned schedule $s$ is
administered at study times $e+s = (e+s_1,\ldots,e+s_k)$.   Denote a patient's time from entry at $e$ to toxicity by $T$,
so that at study time $t$ the patient's observed time to toxicity or last follow-up is
$T^{o}(t) = T$ if  $e+T\leq t$ and $T^{o}(t) = t-e$ if  $e+T>t$.  Defining  $\delta(t)
= I(e+T\leq t)$, the patient's outcome data at study time $t$ are $Y(t) = (T^{o}(t),\delta(t)).$
The probability model is constructed from the patient's hazard of toxicity, $h(u \vert d, \theta)$,
associated with a single administration of dose $d$ given $u$ days previously,
and we denote $H(x| d, \theta) = \int_0^x h(u | d, \theta)\, du.$
Under the assumption that effects of successive administrations of the agent are additive,
the overall hazard of toxicity at study time $t$ for a patient entering at $e$ and treated with $x = (s,d_s)$ is\vspace*{-2pt}
\begin{equation}
\qquad\ \lambda(t |  e, (s,d_s), \theta) = \sum_{j=1}^{k}h\bigl(t-e-s_j |  d(s_j), \theta \bigr),\vspace*{-2pt}
\end{equation}
where $h(u|d,\theta) = 0$ for all $u<0$.
The patient's cumulative hazard of toxicity at study time $t$ is thus\vspace*{-2pt}
\begin{equation}
\qquad\ \Lambda(t | e, (s,d_s), \theta) = \sum_{j=1}^{k} H\bigl(t -e- s_j |   d(s_j), \theta\bigr) ,\vspace*{-2pt}
\end{equation}
and the probability that the patient has not had toxicity by study time $t$ is $\operatorname{Pr}(e+T>t |  e, (s,d_s), \theta)
= \exp\{-\Lambda(t |  e,(s,d_s),\theta\}$.  Thus, $h$ and $H$ are expressed in terms of patient time,
whereas $\lambda$ and $\Lambda$ are expressed in terms of study time.
The probability distribution of $T$ is determined by the particular form of the single administration hazard function $h$.

The model allows each patient's actual $x = (s, d_s)$ received to fall outside the set of $KJ$ treatment
configurations in $\mathcal{X}$, provided that each dose in $d_s$ is an element of $\{d^{(1)},\ldots,d^{(K)}\}.$
In particular, the elements of $d_s$ need not be identical.
This accommodates the possibility that a patient's treatment does not go as planned,
for example, due to interim dose reductions following moderate toxicity or deviations from the planned schedule.
It also allows the possibility that the patient's planned $x$ may be changed before the schedule is completed,
based on other patients' data observed during the patient's therapy.

\begin{figure}[b]

\includegraphics{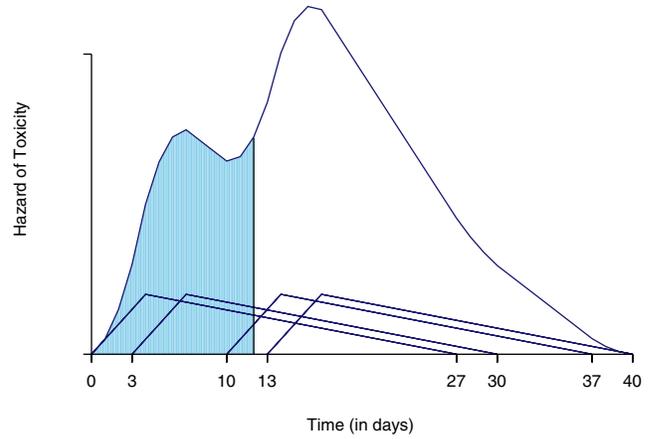}

  \caption{Illustration of triangular component hazards for administrations on days 0, 3,
10 and 13, and the resulting cumulative hazard function.  The shaded region is the cumulative
hazard, $H(12)$, of toxicity by day 12.}\label{fig5}
\end{figure}

At study time $t$, let $x_i(t)$ denote the portion of the $i$th patient's treatment regime $x_i$ that has been administered
by that time and let $\mathcal{D}_t = \{(T_i^{o}(t),\delta_i(t),\break e_i,x_i(t)), i=1,\ldots,n(t) \}$
denote the current data.  The likelihood at study time $t$ is\vspace*{-2pt}
\begin{eqnarray}\label{dslike}
\quad\mathcal{L}(\mathcal{D}_t|\theta) &=&
\prod_{i=1}^{n(t)}\{\lambda(T^{o}_i(t) |  e_i, x_i(t),\theta)\}^{\delta_i(t)}\nonumber
\\[-9pt]\\[-9pt]
&&{}\hspace*{12pt}\cdot\exp\{-\Lambda(T^{o}_i(t) |  e_i, x_i(t), \theta)\}.\nonumber\vspace*{-2pt}
\end{eqnarray}

BTND assume that, for each $j=1,\ldots,J,$
the single-administration hazard function associated with dose $d^{(j)}$ is a triangle, formally
\begin{eqnarray}
h\bigl(u  |  d^{(j)}, \theta_j\bigr) &=& \frac{2\alpha_j u}{(\beta_j+\gamma_j) \beta_j} I(0 \le u \le \beta_j)\nonumber
\\
&&{} +
\frac{2\alpha_j (\beta_j+\gamma_j-u)}{(\beta_j+\gamma_j) \gamma_j}
\\
&&{}\hspace*{10pt}\cdot I(\beta_j < u \le \beta_j+\gamma_j),\nonumber
\end{eqnarray}
where $\theta_j = (\alpha_j,\beta_j,\gamma_j)$, so that  $\theta = (\theta_1,\ldots,\theta_J)$
has $3J$ elements.
The $j$th triangle has base of length $\beta_j+\gamma_j$, area $\alpha_j,$ and maximum height
$h(\beta_j|d^{(j)}, \theta_j) = 2\alpha_j/(\beta_j+\gamma_j)$.  Thus, for $u \geq \beta_j+\gamma_j$,
the cumulative single-administration hazard is $H_j(u|d^{(j)}, \theta_j) = \alpha_j$.
Under this model, since any schedule $(s_1,\ldots,s_k)$ is finite, given $\theta$
and $k$-vector of PADs $d_s = (d^{(j)},\ldots,\break d^{(j)}),$
the cumulative hazard $\Lambda(t |  (s,d_s), \theta)$ has the finite maximum
value $k\alpha_j$.  Consequently,  given $\theta,$ the probability that the patient never experiences toxicity
is $\bar F(t | (s,d_s), \theta) = \exp(-k\alpha_j)$ for all $t > s_k+\beta_j+\gamma_j,$
with the obvious elaboration of the upper limit on $\bar F(t | (s,d_s), \theta)$ if the elements of $d_s$ are not identical.

The triangular form of $h$ may seem to be an oversimplification of a complex phenomenon. In
application, however, it is quite flexible and yields a very robust trial design.
Figure~\ref{fig5} shows the cumulative hazard of toxicity for a patient treated with a fixed PAD according to
the 4-administration schedule $s = (0,3,10,13)$.  The
shaded area represents $H(12)$, the cumulative hazard of toxicity by day 12.  The smoothness of the
curve $H(u)\mbox{ for } 0\leq u \leq 40$ that results from summing four triangles and the fact that the parameters
$(\alpha_j,\beta_j,\gamma_j)$ characterizing the $j$th triangle corresponding to $d^{(j)}$ are estimated
from the accumulating data together provide an intuitive motivation for the model's flexibility and robustness.  This
was borne out by the extensive simulation studies reported by BTND.  In the setting where only schedule is varied
with PAD fixed, Liu and Braun (\citeyear{2009Liu}) have studied the use of a smooth component hazard function, a 2-parameter Weibull,
$h(u|\alpha,\beta) = e^{\beta} \alpha u^{\alpha-1} \exp(-u^{\alpha} e^{\beta})$ for $u>0,$
where $\alpha>0$ and $\beta$ is real-valued. This allows $h$ to be nonmonotone if $\alpha \geq 2$
or decreasing if $0 \leq \alpha < 2$.

\subsection{Trial Conduct}

Given an interval  $[0, t^*]$ large enough to
reliably evaluate $T$ under the longest schedule,
the physician specifies a target $\pi^* = \operatorname{Pr}(T \leq t^*)$.
For brevity, I will temporarily index a patient's assigned treatment $(s^{(k)},d^{(j)})$  by
$(k,j)$, and denote the c.d.f. of $T$ associated with  $x = (k,j)$ by $F_{k,j}(\theta) = \operatorname{Pr}(T\leq t^*|(k,j),\theta)$.
Since $\mathcal{L}(\mathcal{D}_t|\theta)$ and hence the posterior $p(\theta|\mathcal{D}_t)$ change continuously
with $t$ during the trial, necessarily, $x\in\mathcal{X}$ is chosen for each newly enrolled patient,
that is, $c = 1$.  A patient accrued at study time $t$
is assigned the pair $(k,j)\in \mathcal{X}$  minimizing the objective function
$|\mathrm{E}\{F_{k,j}(\theta)|\mathcal{D}_t\} - \pi^*|$, similar to the CRM.
Assignment of $x$ using this criterion is
subject to the following two safety rules.  Given a maximum toxicity probability, $F_{\mathit{max}},$
specified by the physician, the schedule-dose pair $(k,j)$ is {\it acceptable} if $\operatorname{Pr}(F_{k,j}(\theta) > F_{\mathit{max}} |\mathcal{D}_t) < p^*,$
where $p^*$ is a fixed upper cut-off such as 0.80 or 0.90.  If no pair in $\mathcal{X}$ is acceptable,
the  trial is stopped.  This is  similar to the toxicity portion of the acceptability criteria $(\ref{stopET})$ of the
EffTox methodology.  The second safety rule is that escalation from $(k,j)$ is restricted in that no
untried dose-schedule combination may be skipped, specifically the next patient
may be treated at $x = (k+1, j)$, $(k, j+1)$ or $(k+1,j+1)$, but at no higher pair.
There is no such constraint on de-escalation.  While developing this methodology, we initially tried using
the more restrictive constraint that does not allow diagonal escalation, from $(k,j)$
to untried $(k+1,j+1)$, but this  yielded a design with very poor properties.  This is the case essentially because
this constraint makes exploration of the 2-dimensional set of $KJ$ schedule-PAD pairs unfeasible,
and, in fact, it provides no additional measure of safety. While BTND recommended that the first patient be treated at the
safest pair $(k,j) = (1,1)$, in practice, the physician might wish to start at $(1,2)$, $(2,1)$ or $(2,2)$.

It may seem self-evident that this method is greatly superior to any comparable method that fixes schedule
and only varies dose, since an  optimal combination $(s^{(k)},d^{(j)})$ is simply ignored
if the fixed schedule is not~$s^{(k)}$.  The simulations reported by BTND clearly illustrate  this
point.  Currently, however, it still is standard practice in
phase I trials to guess what schedule might be best, possibly based on animal data,
and proceed in humans by varying only dose.
As described by BTND, this new methodology has been used to conduct
an allotx trial of the post-transplant agent 5-azacitidine,
which is thought to kill leukemia cells by reactivating tumor suppressor genes while also
enhancing graft-versus-leukemia effect.

\section{Discussion}

 Each of the designs reviewed here includes one or more more aspects of
treatment or outcome in an early phase trial that are ignored by standard designs.
The price of accommodating such additional complexity is  a much more structured model and method.
This often requires substantially more work for trial design and conduct,
including analysis of historical data, elicitation of priors and design parameters,
development of computer software, carrying out simulations to calibrate design parameters
and establish operating characteristics, and the difficult process of real time data monitoring during trial conduct.
In each case, however, the design provides advantages over standard methods so large that the comparisons
may seem unfair.  Evidently, accounting for both anti-disease effect and toxicity is a good idea,
ignoring covariates is a bad idea, quantifying the clinical importance of different types
and grades of toxicities is a good idea, and ignoring schedule effects is a bad idea.
For example, the optimal treatment pair determined by the design in the 5-azacitidine trial was
(40 mg$/$m$^2$ per administration, 3~cycles), which simply could not have been found using a conventional
dose-finding method that fixes schedule at 1 cycle. While each of the designs relies on a model
with a nontrivial number of parameters, which in turn often requires  elaborate prior specification
and sophisticated numerical methods, the amount of information per patient also is much
greater. The final questions are whether such designs have good properties, which
the computer simulations show they do, and whether they can be implemented in practice,
which has been the case for all of the designs discussed here.

For most early phase trials, conventional methods for determining sample size
based on hypothesis testing or estimation may be of little use.
To determine a planned maximum sample size for a phase I or I/II trial, I ask the physicians
the anticipated accrual rate, which often is a range of values,
the desired maximum trial duration, and cost or other resource limitations, such as the
amount of a specialized agent that feasibly can be produced in the laboratory.
For each of several feasible maximum sample sizes,
I simulate the trial and also compute posterior estimates of important parameters  based
on illustrative data sets.  I then show these results to the physician and ask
him/her to choose a maximum sample size on that basis.  If the largest feasible sample
size does not yield a reasonably reliable design, I~recommend that
the trial not be conducted.  Using this practical approach, in my experience
planned phase~I or I/II sample sizes usually range from 24 to~60.

The most severe difficulties in achieving widespread implementation of outcome-adaptive methods
in early phase clinical trials are computational and sociological.
The first practical requirement is portable, high quality computer software for implementation,
including statistical programs that perform the necessary computations
for specific methods and, ideally, graphical user interfaces that communicate with established databases
and statistical programs to
facilitate  real-time data entry and computation of adaptive decision criteria during the trial.
An ``elephant in the living room'' of
outcome-adaptive methods for clinical trials is that constructing and implementing such information systems
in medical environments is often much
more difficult, expensive and time-consuming than developing a particular statistical method.
Moreover, once such a system is in place, the process of entering the patient data required by an outcome-adaptive
method is time-consuming and potentially error prone.

A natural question is whether one can construct practical designs that address
the problems that arose in the SCT trial described in the \hyperref[sec1]{Introduction}.
Such designs would optimize multiple schedule-dose combinations of several agents used in combination
based on a vector of appropriately chosen efficacy and toxicity outcomes, possibly
accounting for patient covariates and schedule-dose-covariate interactive effects
on the outcomes, while adaptively choosing patient-specific
schedule-dose combinations in real time.  This also would require a decision criterion
based on multiple outcomes, possibly using either efficacy-toxicity trade-offs or numerical
utilities\break
(Houede et al., \citeyear{2010Houede}). Currently, we are working to develop new designs
that include various combinations of these extensions.  In my experience, however, early phase trials are
so complex that most trials cannot be optimally designed until after they already have been carried out,
and a ``one size fits all'' design simply does not exist.

Clinical trials are viewed very differently by the pharmaceutical companies who produce and supply
new agents, by regulatory agencies, by institutional review boards, by administrators who provide
infrastructure and resources for trial conduct, by the
physicians and nurses who actually treat the patients in a trial, and by the patients themselves.
Individuals with decision-making authority in all of these different groups must agree on a trial design
before a trial may be conducted.
Many of these people regard the structure and properties of a particular statistical design as technicalities  too
complicated to understand and at most marginally relevant. Most early phase trials are conducted
using very simple conventional methods that do not require computers and are easy to implement.
Accounting more fully for the complexities of both the \mbox{actual} treatment regimes and the
patients' clinical outcomes is a double edged sword, since the greater safety and reliability that such methods
provide is obtained only by working much harder in both design formulation and trial conduct.
Physicians who understand the advantages of properly constructed outcome-adaptive designs
and want to use such methods are a minority, although they often provide the initial motivation
for developing new statistical designs. Their desire to use outcome-adaptive methods,
and the recent shift in the pharmaceutical community,
at least among statisticians, to embrace all things  ``adaptive'' in clinical trials
seem to be\break harbingers of a different future.  How this may actually translate into practical reality
in the coming years remains to be seen.

The website
\href{http://biostatistics.mdanderson.org/SoftwareDownload}{http://biostatistics.mdanderson.org/}
\href{http://biostatistics.mdanderson.org/SoftwareDownload}{SoftwareDownload}
contains computer programs for implementing the methods described in Sections~\ref{sec2} (ToxFinder),
\ref{sec3.1} and \ref{sec3.2} (EffTox) and \ref{sec5} (Dose Schedule Finder).

\section*{Acknowledgment}

This research was partially supported by NCI Grant 2RO1 CA083932.

\end{document}